\documentclass[pre,twocolumn,superscriptaddress]{revtex4}
\pdfoutput=1
\usepackage{graphicx}
\usepackage{amsmath}
\usepackage{amssymb}
\usepackage{color}
\usepackage{bm}

\definecolor{red}{rgb}{0.75,0,0}
\definecolor{blue}{rgb}{0,0,0.75}
\definecolor{green}{rgb}{0,0.5,0}

\newcommand{\revision}[1]{{#1}}

\begin{document}
\title{Polar Patterns in Active Fluids}

\author{Luca Giomi}
\affiliation{School of Engineering and Applied Sciences, Harvard University, Cambridge, MA 02138,  USA}

\author{M. Cristina Marchetti}
\affiliation{Physics Department and Syracuse Biomaterials Institute, Syracuse University, Syracuse New York, 13244-1130, USA}

\date{\today}

\begin{abstract}
We study the spatio-temporal dynamics of a model of polar active fluid in two dimensions. The system exhibits a transition from an isotropic to a polarized state as a function of density. The uniform polarized state is, however, unstable above a critical value of activity. Upon increasing activity, the active fluids displays increasingly complex patterns, including traveling bands, traveling vortices and chaotic behavior. The advection arising from the particles self-propulsion and unique to polar fluids yields qualitatively new behavior as compared to that obtain in active nematic, with traveling-wave structures. We show that the nonlinear hydrodynamic equations can be mapped onto a simplified diffusion-reaction-convection model, highlighting the connection between the complex dynamics of active system and that of excitable systems.
\end{abstract}

\maketitle

\section{\label{sec:introduction}Introduction}

Bacterial suspensions, \emph{in vitro} mixtures of cytoskeletal filaments and motor proteins, and migrating epithelial cell layers are examples of \emph{active} fluids composed of interacting units that consume energy and collectively generate motion and mechanical stresses.  Due to their elongated shape, active particles can exhibit orientational order at high concentration and have been likened to ``living liquid crystals"~\cite{Gruler1999}. They have been modeled by continuum equations built by modifying the hydrodynamics of liquid crystals  to include nonequilibrium terms that account for the activity of the system \cite{TonerTu1995,TonerRev,SimhaRamaswamySP02,Kruse2004,Kruse2005,SaintillanShelley:2007,SaintillanShelley:2008,EdwardsYeomans2009}, or derived from specific microscopic models~\cite{TBLMCM2003,TBLMCMbook}.

The theoretical and experimental study of active fluids has revealed a wealth of emergent phenomena, including  spontaneously flowing states~\cite{Voituriez06,Kruse2005,Marenduzzo2007,GiomiMarchettiLiverpool:2008}, unconventional rheological properties~\cite{Hatwalne04,TBLMCM06,CatesEtAl:2008,SokolovAranson2009,GiomiTBLMCM-shear2010,Saintillan2010,Rafai2010}, and new spatiotemporal patterns not seen in passive complex fluids~\cite{Fielding2011,Giomi2011}.  Large-scale swirling patterns and ``flocking" have recently been observed in  motility assays consisting of a highly concentrated suspension of actin filaments propelled by myosin molecular motor proteins tethered to a plane~\cite{Schaller2010,Butt2010}. Experiments in bacterial suspensions~\cite{Kaiser2003,Zumsan2007} and simulations of self-propelled hard rods~\cite{Peruani2006,Chate2008,Yang2010,GinelliPeruani2010} have also shown the formation of polar high-density clusters traveling trough a low density background.   In a recent paper one of us and collaborators examined the spatio-temporal dynamics of  active fluids with nematic symmetry in two dimensions, demonstrating the  onset of spatially modulated relaxation oscillations and the close analogy between the dynamics of active fluids and that of  excitable media~\cite{Giomi2011}.

In this paper we consider the spatio-temporal dynamics of an active suspension with polar symmetry. The model is appropriate to describe systems such as bacterial suspensions, where the particles are swimmers, hence  head-tail asymmetric or polar. As a result, the active fluid can order in a macroscopic polar state, characterized by a nonzero value of a vector order parameter ${\bf P}$, describing the mean polarization of the system. The order parameter can be written in terms of a magnitude $P$ and a polar director ${\bf p}$ (with $|{\bf p}|=1$), denoting the direction of spontaneously broken symmetry in the ordered state. While in the nematic state  the director ${\bf n}$ is a headless unit vector because the ordered state  is invariant for ${\bf n}\rightarrow -{\bf n}$, the polar state does not possess this symmetry and the polar director ${\bf p}$ is a true unit vector. As a result, the continuum equation for a polar active fluids contain terms that are forbidden by symmetry in the case of an active nematic. 
These terms describe swimming of the active particles relative to the bulk fluid and yield new self-advection contributions to the continuum equations.

Continuum  descriptions of the collective dynamics of polar active particles have been discussed before in the literature in two cases: for particles moving on a frictional substrate (``dry'' system) and for particles swimming in a fluid (active suspension). The hydrodynamics of the dry system is a continuum theory of the Vicsek model of flocking, consisting of a collection of self-propelled particles moving on a frictional substrate with noisy aligning rules. It was first introduced phenomenologically by Toner and Tu~\cite{TonerTu1995,TonerRev} and recently derived by explicit coarse graining of the microscopic dynamics~\cite{BDG2009,Ihle2011}. In this case the only conserved variable is the density of active particles as momentum is not conserved.  The theory is then formulated entirely in terms of density and polarization fields and the equations lack Galileian invariance due to the presence of  the substrate. The hydrodynamic equations of active suspensions have been written down phenomenologically on the basis of symmetry considerations and also derived from specific microscopic models of a collection of interacting active particles. The equations used here were first obtained by Liverpool and Marchetti for a suspensions of cytoskeletal filaments with crosslinking motor proteins~\cite{TBLMCMbook}. While in this case the filaments may be better described as a ``mutually propelled", rather than ``self-propelled", as the activity resulting in propulsion arises from the forces exchanged among filaments by the active cross-linkers, the form of the hydrodynamic equations does not depend on such microscopic details. 

This work focuses on \revision{pattern formation in polar active suspensions. While the behavior of polar active films in a quasi-1D geometry was discussed before in Ref.~\cite{GiomiMarchettiLiverpool:2008}, here we present a systematic study of a two-dimensional system with periodic boundary conditions. New results include:  the derivation of a ``phase diagram" for the system that unifies the analysis of  instabilities discussed previously in the literature in various contexts (Fig.~\ref{fig:phase-diagram}),  the prediction of the traveling vortices regime shown in Fig.~\ref{fig:traveling-vortices},  the derivation of a low dimensional model that captures the onset of the traveling waves regime, and the prediction that the transitions between the various regimes are discontinuous, allowing for the possibility of coexistence and hysteresis. } 

We begin by identifying the  linear instabilities of the ordered state of polar active fluids, summarized in Fig.~\ref{fig:phase-diagram}. These instabilities have been discussed before in the literature, either in the context of dry systems \emph{or} of active suspensions. Here we   highlight the connection between the behavior of the suspension and the dry limit. To go beyond the linear case, we then integrate numerically the nonlinear hydrodynamic equations for increasing values of activity. In a polar system there exist two classes of active terms. The first class, with strength controlled by a parameter we call $\alpha$, is present in both systems of nematic and polar symmetry and describes active stresses induced by coupling of orientation and flow. The sign of $\alpha$ identifies active systems that generate contractile stresses ($\alpha>0$) versus those that generate tensile stresses ($\alpha<0$). The second class, with strength controlled by a parameter we call $w$, is unique to polar fluids and describes a variety of nonequilibrium advective contributions to the dynamics arising from the particles' self propulsion. Both parameters are ultimately controlled by the activity of the system, as embodied for instance by the rate of adenosine-triphosphate (ATP) consumption in motor-filament suspensions or by the forces exerted by swimmers on the surrounding fluid in a bacterial suspension. For this reason, and to reduce the number of independent parameters, in most of the following we take $w=f\alpha$, with $f$ a numerical factor of order one, and discuss the behavior of the system for increasing values of the activity $\alpha$. Upon increasing $\alpha$ we observe a variety of increasingly complex patterns. A regime that is unique to polar systems consists of bands extending in the direction orthogonal to that of mean order ($x$ direction) and traveling along $x$. Although traveling bands have been seen in dry systems, either via direct simulations of Vicsek models~\cite{Chate2008} or by numerical solution of the continuum theory~\cite{BDG2009,Mishra2010}, there are important qualitative differences between the traveling bands in dry systems and in  suspensions. In dry systems the bands appear to be  solitary waves of ordered regions of high density and polarization traveling in a disordered background. In suspensions, in contrast, the traveling waves  are mainly controlled by coupled oscillations in the flow and the direction of the polarization. They can be understood in terms of a simplified reaction-diffusion-advection model that highlights the crucial role played by active stresses in providing the energy input needed for setting up the pattern formation and by advective terms unique to polar systems in controlling the traveling nature of the pattern. As the activity is further increased, the traveling wave pattern is replaced by more complex spatial structures, including oscillating flows with pairs of traveling vortices, and eventually chaotic behavior. 
 
\revision{Finally, spatial patterns in active polar fluids in two dimensions have also been discussed by Voituriez et al~\cite{Voituriez2006}. These authors include a coupling between splay and density fluctuations that is present in both equilibrium and active polar fluids, but neglect the convective  couplings unique to active polar systems that are responsible for the oscillatory or traveling nature of some of the spatial patterns found in the present work. The modulated flowing phases identified in Ref.~\cite{Voituriez2006} are characterized by a finite steady flow induced by activity, but do not represent traveling patterns. They are ``equilibrium-like" in the sense that they have a corresponding analogue in equilibrium polar systems~\cite{Blankschtein1985}. When polar convective terms are included the modulated patterns are replaced by traveling vortices.}
 
In Sec. \ref{sec:hydrodynamics} we summarize the hydrodynamic equations of a polar active suspension. The homogeneous states of these equations and their linear stability are discussed in Sec. \ref{sec:stability}, where we also show the connection between the instabilities of the suspension and those of a dry system. The results of the linear stability analysis are summarized in the phase diagram shown in Fig.~\ref{fig:phase-diagram}. In Sec. \ref{sec:patterns} we describe the spatio-temporal patterns obtained by numerical solution of the nonlinear hydrodynamic equations in a planar sample in two dimensions. These include traveling bands, traveling vortices and chaotic flow. We also show that the nonlinear equations can be mapped onto a simplified diffusion-reaction-convection model, highlighting the connection between the complex dynamics of active system and that of excitable systems. We conclude with a brief discussion.

\section{\label{sec:hydrodynamics}Hydrodynamics of  polar active suspensions} 

 We consider a suspension of rod-like active particles of length $\ell$ and mass $M$ in a fluid.   We assume that the dynamics of the system can be described by continuum fields consisting of conserved quantities: the concentration $c$ of active particles, the total density  $\rho=\rho_{\rm solvent}+Mc$ of the suspension, assumed constant, and the total momentum density ${\bf g}=\rho{\bf v}$, with ${\bf v}$ the flow velocity. In addition, to describe the possibility of polar order we consider the dynamics of the polarization ${\bf P}$. The concentration and polarization vector of the active particles can be written as~\cite{TBLMCMbook}
\begin{subequations}
\begin{gather}
\label{c}c({\bf r},t)=\langle\sum_n\delta({\bf r}-{\bf r}_n(t))\rangle\;,\\
\label{P}{\bf P}({\bf r},t)=\frac{1}{c({\bf r},t)}\langle\sum_n\hat{\bm\nu}_n(t)\delta({\bf r}-{\bf r}_n(t))\rangle\;,
\end{gather}
\end{subequations}
where ${\bf r}_n(t)$ is the position of the center of mass of the $n$-th swimmer and $\hat{\bm\nu}_n(t)$ is a unit vector along the long axis of the swimmer. We consider here uniaxial swimmers, which are the simplest type of active polar particles.

Although the active suspension is a nonequilibrium system for which a free energy cannot be defined, it is convenient to write the hydrodynamic equations in terms of a ``free energy'' given by
\begin{multline}\label{eq:free-energy}
F = \int d{\bf r}\,\bigl[\tfrac{1}{2}a_{2}|{\bf P}|^{2}+\tfrac{1}{4}a_{4}|{\bf P}|^{4}\\
+\tfrac{1}{2}K\partial_{i}P_{j}\partial_{i}P_{j}-\tilde{w}_{3}\,(\delta c/c_{0})\bm{\nabla}\cdot{\bf P}\bigr]\;,
\end{multline}
with $\delta c=c-c_0$ the deviation of the concentration from its mean value, $c_0$. The first two terms on the right hand side of Eq. \eqref{eq:free-energy} determine the homogeneous steady state of the system.  The parameter $a_{2}\sim c^*-c$ changes sign at a critical concentration $c^*$, while $a_{4}>0$. The third term describes the energy cost of spatially inhomogeneous deformations of the order parameters, with $K$ a stiffness. For simplicity we have used a one-elastic constant approximation and equated the elastic constants for splay and bend deformations. Finally, the last term describes a coupling between concentration and splay deformation that is only allowed in a polar fluid. We have neglected polar couplings between splay and fluctuations in the magnitude of the order parameter  $\sim |{\bf P}|^2\bm\nabla\cdot{\bf P}$ that yield advective-type terms and pressure corrections  in the equation of motion for the order parameter. 

The dynamics of the concentration of active particles is governed by a convection-diffusion equation
\begin{equation}
\label{c-eq}
\partial_t c+\bm\nabla\cdot c\big({\bf v}+w_1{\bf P}\big)=\bm{\nabla}\cdot(\bm{D}\bm{\nabla} c)\;,
\end{equation}
where $D_{ij}=D_{0}\delta_{ij}+D_{1}P_{i}P_{j}$ is the anisotropic diffusion tensor. The equation for the polarization is given by:
\begin{equation}
\left[\partial_t+({\bf v}+w_2 {\bf P})\cdot\bm\nabla\right]P_i+\omega_{ij}P_j=
\lambda u_{ij}P_j+\gamma^{-1}\,h_i
\;,\label{P-eq}
\end{equation}
where ${\bf h}=-\delta F/\delta {\bf P}$ plays the role of the nematic molecular field, with:
\begin{equation} 
{\bf h}=-\left(a_2+a_4|{\bf P}|^2\right)\,{\bf P}+K\nabla^2 {\bf P}-\tilde{w}_{3}\,\frac{\bm{\nabla} c}{c}\;.
\;,\label{h}
\end{equation}
Also, $\gamma$ is a friction and $\lambda$ a reactive parameter that controls the coupling between orientation and flow. As in a nematic liquid crystal, order parameter variations are coupled to flow gradients, described by the antisymmetric and symmetric parts of the rate of strain tensor $u_{ij}=(\partial_{i}v_{j}+\partial_{j}v_{i})/2$ and $\omega_{ij}=(\partial_{i}v_{j}-\partial_{j}v_{i})/2$.

Finally,  $w_1$ and $w_2$ are characteristic velocities (taken independent of $c$), proportional to the self-propulsion speed of the active particles, and describe self-advection  of the active particles relative to the fluid. These terms are unique to a polar active fluid. The evolution of the flow velocity is controlled by the Navier-Stokes equation
\begin{equation}
\rho\left(\partial_t+{\bf v}\cdot\bm\nabla\right)v_i=\partial_j\sigma_{ij}\;,
\label{NS}
\end{equation}
with $\bm\nabla\cdot{\bf v}=0$, as required by incompressibility. The  stress tensor $\sigma_{ij}$ can be written as
\begin{equation}
\sigma_{ij}=2\eta u_{ij}+\sigma_{ij}^r+\sigma_{ij}^a\;.
\label{sigma}
\end{equation}
The first term on the right hand side of Eq. (\ref{sigma}) is the dissipative part and we have used an isotropic-viscosity approximation. The second term is the reversible part, given by
\begin{equation}
\sigma_{ij}^r=-\Pi\delta_{ij}-\frac{\lambda}{2}(P_ih_j+P_jh_i)+\frac{1}{2}(P_ih_j-P_jh_i)-\bar\lambda\delta_{ij}
{\bf P}\cdot{\bf h}\;,\label{sigma-r}
\end{equation}
with $\Pi$ the pressure and $\bar\lambda$ another reactive parameter. In a nonequilibrium system the coefficients of the various terms in the constitutive equation for the stress tensor are not in general related to those in the molecular field ${\bf h}$, as is instead required in equilibrium. Here we have taken them to be the same. This simplifying assumption does not change the structure of the equation, but only serves to limit the number of independent parameters in the equations. It is a direct consequence of having written the reactive part of the fluxes in terms of the derivative of a free energy. Finally, the last term in Eq.~\eqref{sigma} is the active contribution. It can be written as the sum of two terms of different symmetry,
$\sigma_{ij}^a=\sigma_{ij}^\alpha+\sigma_{ij}^\beta$, with
\begin{subequations}
\begin{gather}
\sigma_{ij}^\alpha=\alpha c(P_iP_i+\frac{1}{2}\delta_{ij}P^2)\label{sigmaa}\;,\\
\sigma_{ij}^\beta=\beta c(\partial_iP_j+\partial_jP_i+\delta_{ij}\bm\nabla\cdot{\bf P})\;,\label{sigmab}
\end{gather}
\end{subequations}
where $\sigma_{ij}^\alpha$ describes tensile/contractile active stresses present in both nematic and polar fluid, while $\sigma_{ij}^\beta$ is unique to polar fluids and describes active stresses due to self-advection and treadmilling \cite{defnote}. Note that $\alpha>0$ corresponds to contractile stresses as generated by pullers or by motor-filament systems, while $\alpha<0$ describes tensile stresses as generated by most swimming bacteria, such as E. coli~\cite{alpha-sign}. In the following we will ignore the polar contribution to the active stress and simply incorporate the effect of polarity via the advective terms proportional to $w_i$. We will further assume $w_1=w_2=2\tilde{w}_3/\gamma=w$. We  make the system dimensionless by scaling all lengths with the length $\ell$ of the active particles, scaling time with the relaxation time of the director $\tau_p=\ell\gamma/K$ and scaling stress using the elastic stress $\sigma=K/\ell^2$.

\section{\label{sec:stability}Homogeneous steady states and their stability} 

There are two homogeneous, steady state solutions of the hydrodynamic equations, both with constant density $c_0$. For $c_0<c^*$, or  $a_{20}\equiv a_2(c_0)>0$, the homogeneous solution  is the isotropic state, with  ${\bf P}=0$. For $c_0>c^*$, or $a_{20}<0$, rotational symmetry is spontaneously broken and there is an ordered polarized state, with  $|{\bf P}|=P_0=\sqrt{-a_{20}/a_{40}}$. In the numerical work described below we have used the following form for the coefficients $a_2$ and $a_4$:
\[
a_{2}=c^{*}-c\,,\qquad
a_{4}=c^{*}+c\,.
\]
This choice ensures that $P_0\sim \sqrt{c_0-c^*}$ for $c_0\rightarrow c^*_+$ and $P_0\sim 1$ for $c_0\gg c^*$.

Next, we examine the linear stability of the homogeneous states against spatially inhomogeneous fluctuations in the continuum fields. The isotropic state is always linearly stable and will not be discussed further. In order to ensure the incompressibility condition, it is convenient to rewrite the Navier-Stokes equation in terms of vorticity and stream function $\psi$ by expressing:
\[
v_{x} = \partial_{y}\psi\qquad v_{y}=-\partial_{x}\psi\;.
\] 
The incompressibility condition is then automatically satisfied. The vorticity field is given by
\begin{equation}\label{eq:vorticity}
\omega = 2\omega_{xy} = \partial_{x}v_{y}-\partial_{y}v_{x}\,,
\end{equation}
and the stream-function $\psi$ is related to the vorticity $\omega$ through a Poisson equation: 
\begin{equation}
\label{Poisson}
\nabla^{2}\psi = -\omega\,.
\end{equation}
The Navier-Stokes equation can be written in terms of $\omega$ and $\psi$ as
\begin{equation}
\label{NS-2}
\partial_{t}\omega = \eta\nabla^{2}\omega+\partial_{x}^{2}\tau_{yx}+\partial_{xy}\tau_{yy}-\partial_{yx}\tau_{xx}-\partial_{y}^{2}\tau_{xy}\;,
\end{equation}
where we have separated out the viscous part of the stress tensor and defined $\tau_{ij}$ as
\begin{equation}
\label{tau}
\tau_{ij}=\sigma_{ij}^r+\sigma_{ij}^a-\frac12\delta_{ij}\left( \sigma_{kk}^r+\sigma_{kk}^a\right)\;.
\end{equation}

\begin{figure}[t]
\centering
\includegraphics[scale=0.7]{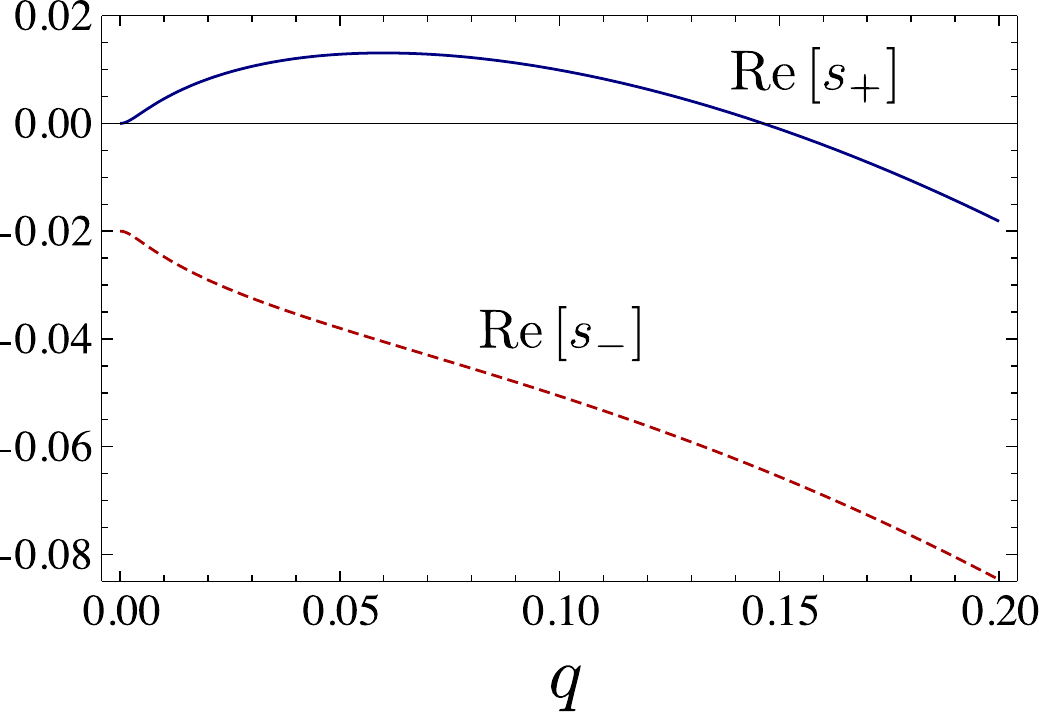}
\caption{\label{fig:density-mode}Real part of the eigenvalues $s_{\pm}$ of the matrix $\bm{a}_{10}$ displaying the coupled instability of $(\delta c,\delta P_x)$ for $w=0.5$ and $c_0=0.01$.}
\end{figure}

To analyze the stability of the polarized state, we choose the  $x$ axis along the direction of order and compactify the notation by introducing a vector 
$\bm{\varphi}=\{c,P_{x},P_{y},\omega\}$.
The homogeneous polarized state is characterized by $\bm{\varphi}^{(0)}=\{c_{0},P_{0},0,0\}$. We then introduce the fluctuations from the homogeneous values by letting
\[
\bm{\varphi}({\bf x}, t) = \bm{\varphi}^{(0)}+\epsilon\,\bm{\varphi}^{(1)}({\bf x},t)
\]
and consider the dynamics of fluctuations for  $\epsilon\ll 1$. To enforce periodic boundary conditions on a square domain, we look for solutions of the form
\[
\bm{\varphi}^{(1)}({\bf x},t) = \sum_{n=-\infty}^{\infty}\sum_{m=-\infty}^{\infty}\bm{\varphi}_{nm}(t)\,e^{i(q_nx+q_my)}\;,
\]
where $q_{n}= 2\pi n /L$ and $q_{m}= 2\pi m/L$. The Fourier components of the stream-function are related to those of the vorticity by:
\[
\psi_{nm} = \frac{\omega_{nm}}{q_n^{2}+q_m^{2}}\;.
\]
With this choice the linearized hydrodynamic equations reduce to a set of coupled linear ordinary differential equations for the Fourier modes $\bm{\varphi}_{nm}$, given by
\begin{equation}
\label{ODE}
\partial_{t}\bm{\varphi}_{nm} = \bm{A}_{nm}\bm{\varphi}_{nm}\;.
\end{equation}
The matrix $\bm{A}_{nm}$ can be expressed in the following block-form
\begin{equation}
\label{Anm}
\bm{A}_{nm} =
\left(
\begin{array}{cc}
 \bm{a}_{nm} & \bm{b}_{nm} \\
 \bm{c}_{nm} & \bm{d}_{nm}	
\end{array}
\right)\;.
\end{equation}
The explicit expression for the matrix $\bm{A}_{nm}$ is given in Appendix \ref{sec:AppendixA}. In order for the homogeneous state $\bm{\varphi}_{0}$ to be stable, the real part of the eigenvalues of the matrix $\bm{A}_{nm}$ must be negative. It is useful to examine the behavior of the eigenvalues for  $(n,m)$ modes $(1,0)$ and $(0,1)$, corresponding to  wavevectors longitudinal and transverse to the direction of mean order, respectively. In both cases $(\delta c, P_x)$ fluctuations decouple from $(P_y,\omega)$ fluctuations and the dynamics can be examined by solving two independent $2\times 2$ eigenvalue problems.  

\begin{figure}[t]
\centering
\includegraphics[width=0.9\columnwidth]{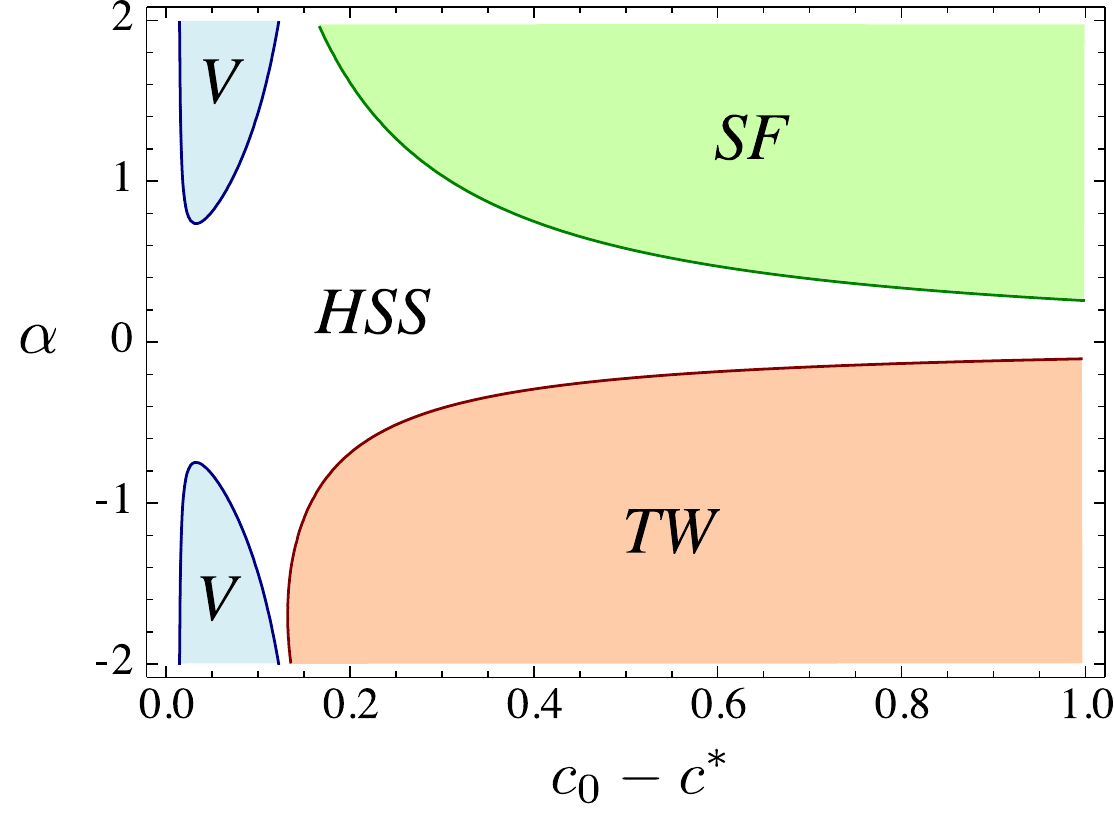}
\caption{\label{fig:phase-diagram} (color online) Phase-diagram in the plane $(c_{0}-c^{*},\alpha)$ for $w=\alpha$ and $q=0.2$, $\lambda=0.5$, $\eta=D_{0}=D_{1}=1$. The labeled regions correspond to the homogeneous steady state ({\em HSS}), the spontaneous flow regime ({\em SF}), the traveling waves regime ({\em TW}) and the region of longitudinal traveling waves of Vicsek-type ({\em V}).}
\end{figure}

We first consider  the $(1,0)$ or longitudinal modes. The matrix  ${\bm b}_{10}=0$, hence ${\bm A}_{10}$ is  block diagonal. The coupled dynamics of $ c$ and $P_x$ fluctuations (note that $\delta P_x$ describes fluctuations in the magnitude of the order parameter)  is controlled by the eigenvalues of the $2\times 2$ matrix $\bm{a}_{10}$, which does not depend on the active parameter $\alpha$, but only on the advective velocity $w$.
The $(10)$ eigenvalues are given by
\begin{multline}\label{density-eigenvalues}
s_{\pm}
=-\frac{1}{2}(C+2iqwP_0)\\
\pm\frac{1}{2}\sqrt{C^2-2w^2q^2-8iqwP_0\frac{c_0c^*}{c_0+c^*}}\;,
\end{multline}
with: 
\[
C=2|c_{0}-c^{*}|+(K+D)q^2\,,
\]
$q=2\pi/L$ and $D_{0}=D_{1}=D$. The fluctuations of concentration and order parameter magnitude in the longitudinal $(1,0)$ exhibit unstable growth for arbitrarily small values of $w$ as the mean field order-disorder transition is approached from above, i.e.,  $c_0\rightarrow c^{*+}$. The real part of the eigenvalues of the matrix $\bm{ a}_{10}$ is shown in Fig.~\ref{fig:density-mode}.  This instability does not couple to the flow and is precisely the instability obtained  in Refs.~\cite{BDG2009,Mishra2010}  for dry systems consisting of self-propelled particles with aligning interactions moving on a frictional substrate. It occurs for both tensile and contractile systems, but it is unique to polar active fluids. The instability only occurs in a narrow range of densities above the putative mean-field continuous transition point $c^*$ between polarized and isotropic states, as shown in the stability phase diagram of Fig.~\ref{fig:phase-diagram}. It has been argued that it may be associated with the renormalization of the order of the transition, which has been shown to be discontinuous in recent simulations of Vicsek-type models~\cite{Chate2008}. We note that the magnitude of the order parameter $P_x$ is not a hydrodynamic variable in the strict sense  as its decay rate has the finite value $-2|a_{20}|$ at large wavelengths. At the transition, however,  $a_{20}= 0$ and fluctuations in $P_x$ become long-lived, driving the instability. In the absence of noise, the numerical solution of continuum equations obtained by coarse-graining the Vicsek model has yielded in this region of the parameters a spatially inhomogeneous state consisting of solitary waves or bands of ordered regions traveling in a disordered background~\cite{BDG2009}.  We refer to the region of parameters where this linear instability occurs as longitudinal traveling Vicsek-type waves ($V$) regime.

The coupled dynamics of fluctuations in the transverse component $P_y$ of the order parameter, describing director fluctuations, and vorticity $\omega$ yields additional instabilities, both in the $(1,0)$ and $(0,1)$ directions, controlled by the $\bm{d}_{nm}$ matrix. The first unstable modes are the longitudinal mode $(1,0)$ and the transverse mode $(0,1)$. These modes becomes unstable for negative and positive values of the combination $\alpha(1-\lambda)$, respectively. These instabilities arise from the coupling of director deformations  and flow and occur even for $w=0$, corresponding to systems with nematic symmetry.   For $\alpha(1-\lambda)>0$, transverse modes $(0,1)$ go unstable, corresponding to coupled splay deformations of the director and fluctuations of the $v_y$ velocity. For $\lambda<1$, the instability occurs above a positive critical value of $\alpha$ given by 
\begin{equation}
\label{eq:alpha01}
\alpha_{01} = \frac{q^{2}[4\eta+P_{0}^{2}(1-\lambda)^{2}]}{2c_{0}P_{0}^{2}(1-\lambda)}\;.
\end{equation}
More precisely  this instability occurs for tumbling ($|\lambda|<1$) or disk-like ($\lambda<-1$) pullers for $\alpha>\alpha_{01}>0$ and for rod-like ($\lambda>1$) pushers for $\alpha<\alpha_{01}<0$. The interplay between the parameter $\lambda$ controlling the flow alignment properties of an orientationally ordered suspension and the activity $\alpha$ in controlling the instabilities of active fluids was discussed in Ref.~\cite{GiomiMarchettiLiverpool:2008} and we refer to that work for details. The critical value $\alpha_{01}$ of transverse or bend instability does not depend on $w$ and is identical to the value obtained for active nematic, with the replacement  $P_0\rightarrow S$, where $S$ is the magnitude of the nematic order parameter. This instability  has been obtained before in the literature~\cite{SimhaRamaswamySP02} and is commonly referred to as the generic instability.  For $\alpha(1+\lambda)<0$, corresponding to tensile active stresses as generated by pushers, such as E. coli, longitudinal modes $(1,0)$ go unstable, corresponding to coupled bend deformations of the director and fluctuations in the $v_x$ velocity. This instability occurs for $\lambda>-1$ and $\alpha<\alpha_{10}<0$ and for $\lambda<-1$ and $\alpha>\alpha_{10}>0$, with 
\begin{equation}\label{eq:alpha10}
\alpha_{10} = -\frac{4\eta q^{2}(1+\eta)^{2}+P_{0}^{2}[4\eta w^{2}+q^{2}(1+\eta)^{2}(1+\lambda)^{2}]}{2c_{0}P_{0}^{2}(1+\lambda)(1+\eta)^{2}} \\[10pt]
\end{equation}
Notice that the critical value $\alpha_{10}$ depends on $w$ (a finite $w$ in this case actually suppresses the instability). When $w=0$, this coincides with the generic instability discussed in \cite{SimhaRamaswamySP02} for pushers and active nematic. For finite $w$, however, the modes that goes unstable is a propagating wave, suggesting that the system may support traveling waves solutions in this region of parameter. This region is labelled at {\em TW} in Fig.~\ref{fig:phase-diagram}. Finally, the splay/bend instabilities obtained here are simply the 2D generalization of the instability to a spontaneously flowing state discussed in Refs.~\cite{Voituriez06} and \cite{GiomiMarchettiLiverpool:2008} for nematic and polar active fluids, respectively, in a quasi-one-dimensional strip geometry. Their origin lies in the long-range nature of the hydrodynamic interactions between active swimming particles, as shown in Ref.~\cite{BaskaranMarchetti:2009}. Some additional detail on the eigenvalues of the $\bm{d}_{nm}$ matrix can be found in Appendix \ref{sec:AppendixB}.

\subsection{\label{sec:dry}Relation to Dry Systems}
For completeness we compare the behavior of the active suspension considered here to that of active particles on a frictional substrate, referred to so far as dry systems. In this case the only conserved field is the concentration of active particles and the continuum equations are written solely in terms of concentration and polarization fields. The momentum of the system is not conserved and there is no equation for the flow velocity~\cite{TonerTu-note}. The low density longitudinal instability of the ordered state obtained near the mean-field order order-disorder transition ($c_0\rightarrow c^{*+}$) is unique to polar active fluids with aligning interactions, corresponding for instance to coarse-grained versions of the Vicsek model, and does not couple to the flow velocity. It is present in both dry systems and suspensions. It was discussed for the case polar particles on a substrate in Refs.~\cite{BDG2009} and \cite{Mishra2010}. 

In contrast, the splay and bend instabilities obtained at high density are due curvature-induced currents arising from the coupling of director deformation and flow. One can obtain the limit of dry systems from our general equations for a suspension by adding a drag $-\zeta {\bf v}$ to the right hand side of Eq.~\eqref{NS}, describing the coupling frictional coupling to a substrate. In the limit of large friction $\zeta$, one can then neglect inertial terms and write the solution of Eq.~\eqref{NS} to lowest order in the gradients as
\begin{equation}
\label{v-damped}
v_i\simeq\frac{1}{\zeta}\left(-\partial_i\Pi+\partial_j\sigma_{ij}^a\right)\;,
\end{equation}
where the active stress $\sigma_{ij}^a$ is given in Eq.~\eqref{sigmaa} and we must satisfy $\bm\nabla\cdot{\bf v}=0$. If Eq.~\eqref{v-damped} is used to eliminate ${\bf v}$ from the equations for concentration and polarization, one then recovers both the splay and bend instabilities. Note that when Eq.~\eqref{v-damped} is used to eliminate the flow velocity in favor of polarization and density the terms responsible for the splay and bend instabilities are of order $(\nabla^2P^3)$. These terms have been neglected in the continuum models of polar fluids on a substrate discussed in the literature, which explain why such instabilities were not obtained to linear order~\cite{Mishra-splay-note}. When considering the full nonlinear equations similar effects are, however, provided by other nonlinearities.

\begin{figure}[t]
\centering
\includegraphics[width=1\columnwidth]{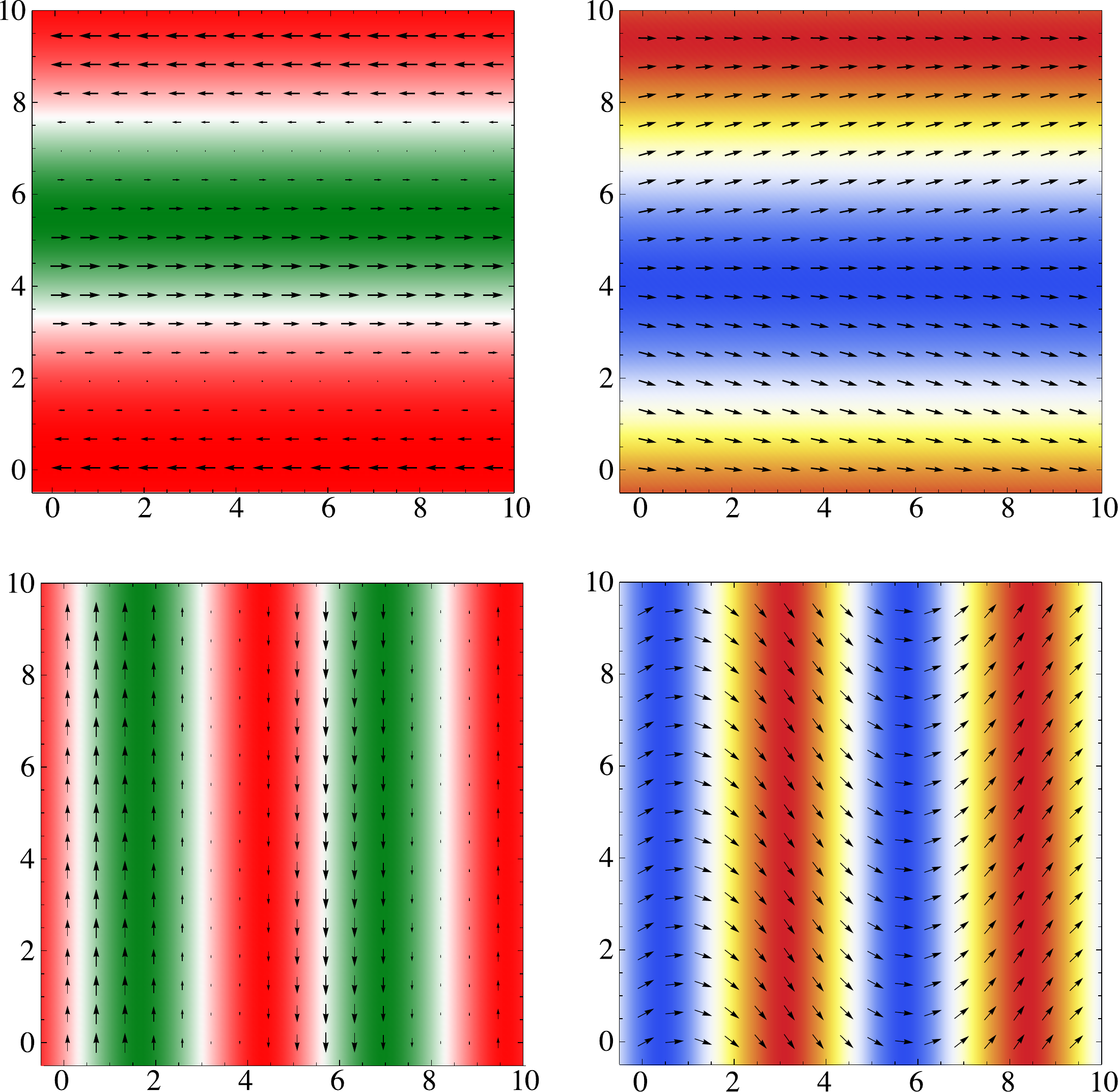}
\caption{\label{fig:traveling-waves}(color online) The velocity field (left) and the polarization direction (right) superimposed to a density plot of the concentration and the polarization magnitude for $\alpha=1.5$ (top) and $\alpha=-2.1$ (bottom). The colors indicate regions of large (red) and small (green) density and large (red) and small (blue) polarization. For positive values of $\alpha$ the homogeneous state undergoes a splay instability that leads to the formation of two stationary bands. For negative $\alpha$ a bending instability results in the appearance of vertical bands traveling at speed $v_{0}\sim w\sim\alpha$ (hence from right to left).}	
\end{figure}

\section{\label{sec:patterns}Nonlinear Spatio-Temporal Patterns}

In this section we report some results obtained from a numerical solution of Eqs. \eqref{c-eq}, \eqref{P-eq} and \eqref{NS}. To explore different regimes at increasing level of activity we set $w=0.1\alpha$ and use $\alpha$ as a variable parameter. Eqs. \eqref{c-eq}, \eqref{P-eq} and \eqref{NS} are then integrated using a vorticity/stream-function finite difference scheme on a collocated grid of lattice spacing $\Delta x = \Delta y = 0.078$. The time integration is performed via a fourth order Runge-Kutta method with time step $\Delta t = 10^{-3}$. As mentioned in Sec. \ref{sec:stability} , the vorticity/stream-function method requires one to solve a Poisson equation at each time step in order to calculate the components of the flow velocity. This was efficiently preformed with a $V$-cycle multigrid algorithm. As initial configurations we consider a homogeneous system whose director field was aligned along the $x$ axis subject to a small random perturbation in density and orientation. Thus $c=c_{0}+\epsilon$, $\theta=\epsilon$, $P=\sqrt{(c-c^{*})/(c+c^{*})}$ and $v_{x}=v_{y}=0$, where $\epsilon$ is a random number of zero mean and variance $\langle \epsilon^{2} \rangle = 10^{-2}$. The equations where then integrated from $t=0$ to $t=10^{3}$, corresponding to $10^{6}$ time steps. For all the numerical calculations described in this section we used $\eta=c^{*}=D_{0}=D_{1}=1$, $\lambda=0.1$ and $L=10$, in the units described in Sec. \ref{sec:hydrodynamics} . Upon varying the activity parameter $\alpha$ five different regimes have been observed: 1) homogeneous stationary state; 2) spontaneous steady-flow; 3) oscillating flow with orthogonal traveling bands; 4) oscillating flow with pairs of traveling vortices and 5) chaotic flow. These regimes are described below in more detail.

\subsection{\label{sec:homogeneous}Homogeneous State and Spontaneous Steady Flow}

For small values of $|\alpha|$, the system quickly relaxes to a homogeneous state with the polarization vector aligned along the $x$ direction. The concentration is equal to the initial concentration $c_{0}$ and the polar order parameter equates its equilibrium value $P_{o}=\sqrt{(c_{0}-c^{*})/(c_{0}+c^{*})}$. Upon raising $\alpha$ above the critical value \eqref{eq:alpha01} , the homogeneous state becomes unstable to a non-homogeneous flowing state (top of Fig. \ref{fig:traveling-waves}). The structure of this state is the same of active nematic fluids and consists of two bands flowing in opposite directions \cite{Giomi2011}. The direction of the streamlines is dictated by the initial conditions which, in this case, favor a flow in the $x$ direction.  This solution has been intensively discussed in the literature of active matter (see for instance \cite{Voituriez06,Marenduzzo2007,GiomiMarchettiLiverpool:2008}) and will not be commented here any further. 

\subsection{\label{sec:travelingwaves}Traveling Waves}

\begin{figure}
\centering
\includegraphics[width=1\columnwidth]{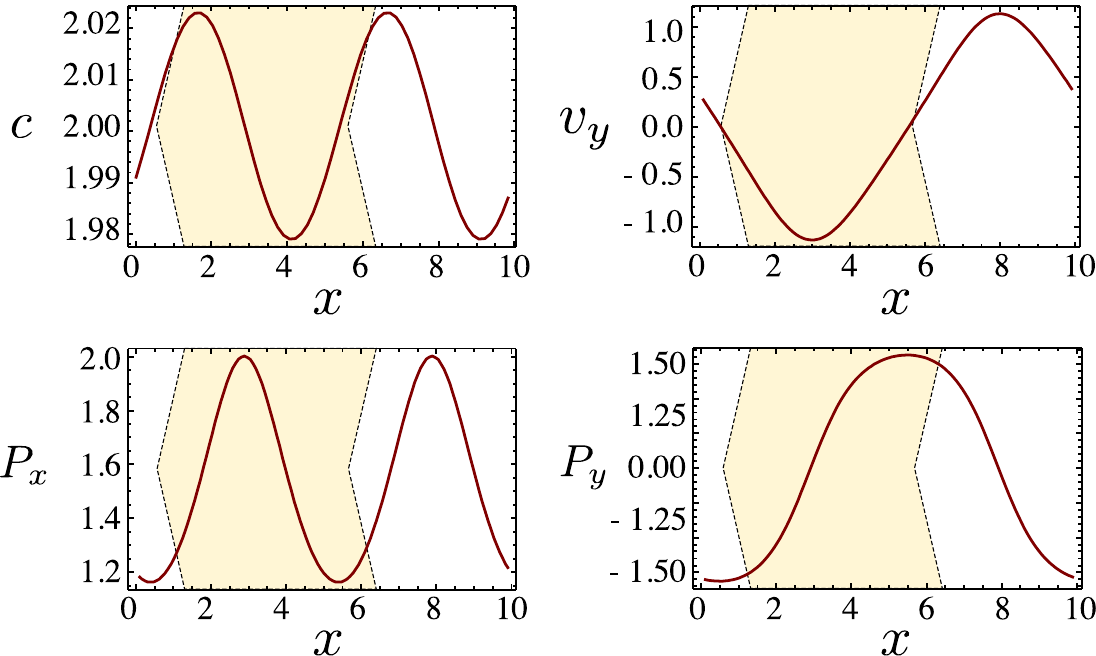}	
\caption{\label{fig:spatial-waves}(color online) The quantities $c$, $v_{y}$, $P_{x}$ and $P_{y}$ along the $x$ axis in the traveling waves regime for $\alpha=-2.1$. The yellow region indicate the band visible in the bottom panel of Fig. \ref{fig:traveling-waves}. The slant of the yellow region indicate the direction of motion of the band.}
\end{figure}

\begin{figure}
\centering
\includegraphics[width=1\columnwidth]{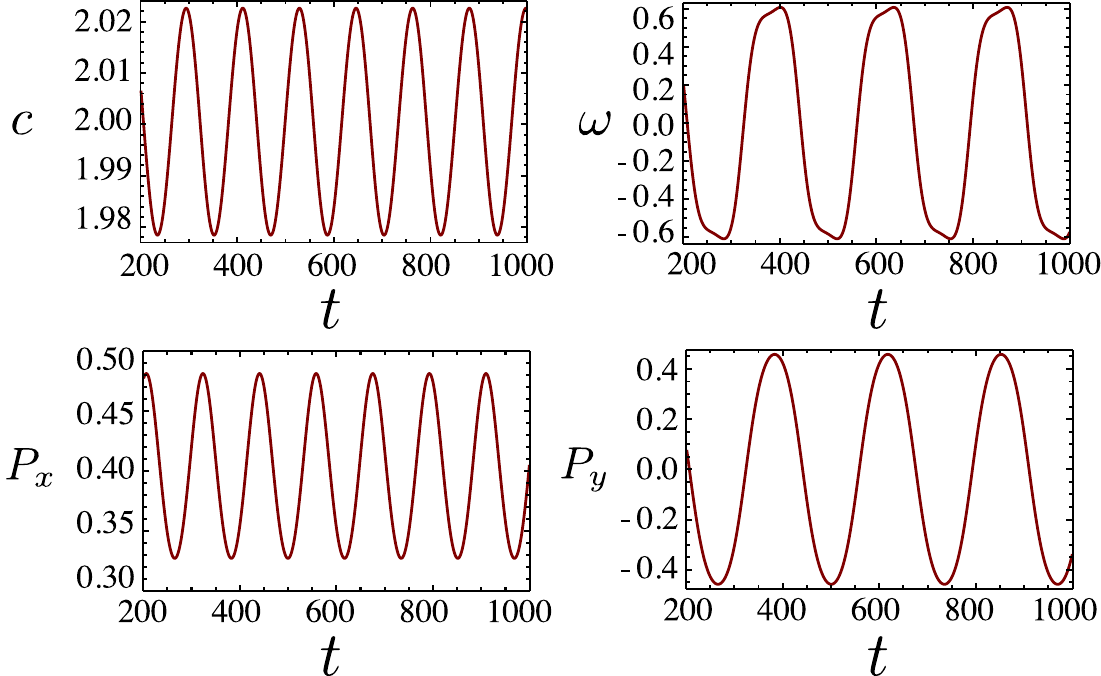}	
\caption{\label{fig:trajectory1}(color online) The quantities $c$, $\omega$, $P_{x}$ and $P_{y}$ at the point $x=y=L/2$ as a function of time in the traveling waves regime for $\alpha=-2.1$.}
\end{figure}

For negative values of $\alpha$ larger in magnitude than the critical value \eqref{eq:alpha10} ($\alpha\approx-1.19$ for our choice of parameters) traveling waves form in the system. As shown in Fig. \ref{fig:traveling-waves} the waves consist of a shear-flow along the $y$ direction, that translates at constant speed along $x$: $v_{x}=0$ and $v_{y}=v_{y}(x-v_{0}t)$ with $v_{0}$ the wave velocity. The concentration $c$ and the polarization field $P$ follows the flow, thus forming traveling bands across the $x$ direction. The polarization direction ${\bf p}$, on the other hand, oscillate while keeping the mean orientation orthogonal to the bands.

The physical origin of the traveling waves in this model of active polar fluids relies on the fact that the polarization field is simultaneously a broken-symmetry variable, a mechanical stress and a physical velocity. Being a broken-symmetry variable, the polarization relaxes toward a non-zero value in the characteristic time scale $\tau_{\rm p}$ (see Sec. \ref{sec:hydrodynamics}). Correspondingly, shear stress $\sigma_{xy}^{a}\sim\alpha P_{x}P_{y}$ is injected in the system at rate $\tau_{\rm a}^{-1}\sim\alpha$. When $\tau_{\rm a}\approx\tau_{\rm p}$ this residual stress is accommodated in the passive structures leading to an elastic distortion and flow as discussed in Sec. \ref{sec:stability}. Unlike nematic fluids, here the existence of a direction of broken-symmetry results in a directed motion of particles with velocity ${\bf v}^{a}\sim w {\bf P}$, hence the pattern produced by the interplay between active stress, liquid crystalline elasticity and flow is advected along ${\bf P}$ resulting in the traveling waves described above. Note that the pattern generated by the splay instability (top of Fig. \ref{fig:traveling-waves}), is not accompanied by the formation of traveling bands since the system is translational invariant along the polarization direction, hence there is no advective transport: i.e. ${\bf P}\cdot\bm{\nabla} = P_{x}\partial_{x}=0$.

To elucidate the properties of this solution, it is useful to reduce the hydrodynamic equations to a tractable form. In the following we will see that under drastic but reasonable simplifications, the hydrodynamic equations presented in Sec. \ref{sec:hydrodynamics} map into a reaction-diffusion-advection system that admits traveling waves solutions. Diffusion-reaction systems have played a fundamental role in the study of biological waves and pattern formation, especially in the context of morphogenesis since the seminal 1952 paper by Turing \cite{Turing:1952}. 

\begin{figure}
\centering
\includegraphics[width=0.75\columnwidth]{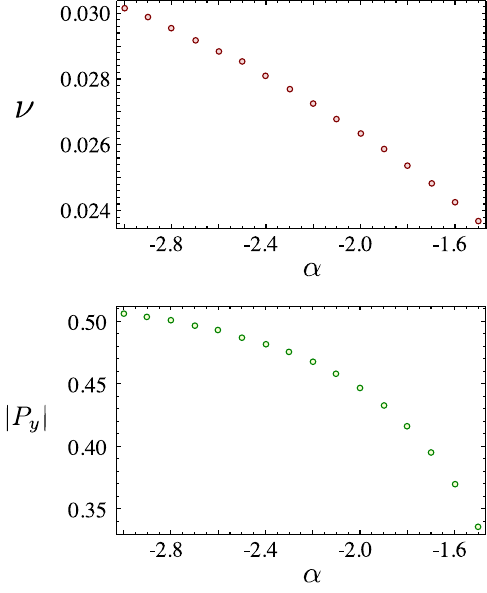}	
\caption{\label{fig:frequency}(color online) The frequency $\nu$ of the waves and the magnitude of $P_{y}$ as a function of $\alpha$ in the traveling waves regime.} 
\end{figure}

As a starting point we notice that at the onset of the traveling bands regime, the variations in $c$ and $P_{x}$ are only slight with respect to those in $\omega$ and $P_{y}$ (see Figs. \ref{fig:traveling-waves} and \ref{fig:trajectory1}). This suggests that the dynamics of the traveling bands is dominated by the latter two fields, \revision{as also indicated by the linear stability analysis.}  Thus we approximate $c$ and $P_{x}$ as constants and focus our attention on the equations for $\omega$ and $P_{y}$:
\[
[\partial_{t}+({\bf v}+w{\bf P})\cdot\bm{\nabla}]P_{y} = \lambda (u_{yx}P_{x}+u_{yy}P_{y})-\omega_{yx}P_{x}+h_{y}\;.
\]
Next, we notice that $v_{x}=0$ and that there is no dependence on the $y$ coordinate. Incompressibility ensures then: $u_{xx}=u_{yy}=0$. Morover $u_{xy}=u_{yx}=\omega_{xy}=-\omega_{yx}=\omega/2$. The previous expression and Eq. \eqref{NS-2} simplify then to:
\begin{subequations} 
\begin{gather}
\partial_{t}P_{y}=\frac{1}{2}P_{x}(1+\lambda)\omega-wP_{x}\partial_{x}P_{y}\notag\\-[a_{2}+a_{4}(P_{x}^{2}+P_{y}^{2})]P_{y}+\partial_{x}^{2}P_{y}\;,\\[10pt]
\partial_{t}\omega = \eta\partial_{x}^{2}\omega+\partial_{x}^{2}\tau_{yx}\;.
\end{gather}
\end{subequations}
Without loss of generality, we take $a_{4}=1$ and $a_{2}+a_{4}P_{x}^{2}=-a$. Now, in order to understand the origin of the traveling bands, one does not need to consider the full stress tensor $\tau_{yx}$, but only the active contribution $\tau_{yx}=\alpha P_{x}P_{y}$, which is responsible for injecting shear stress inside the system at the rate $\alpha$. Thus, taking for simplicity $w=\alpha$ and transforming $\alpha\rightarrow\alpha P_{x}$ we arrive to the following set of equations:
\begin{subequations}\label{eq:simple} 
\begin{gather}
\dot{P}+\alpha P' = (a-P^{2})P+P'' + b\,\omega\;, \\[7pt]
\dot{\omega} = \eta\omega''+\alpha P''\;,
\end{gather}
\end{subequations}
where the dot and the prime indicate respectively the time and space derivative, $b=P_{x}(1+\lambda)/2$ and we have renamed $P=P_{y}$ for shortness. Eqs. \eqref{eq:simple} have the classical form of a diffusion-reaction-advection system. The cubic nonlinearity is the trademark of excitable dynamical systems and, as discussed in Ref. \cite{Giomi2011}, gives the active fluid the behavior of an excitable medium. The advection term $\alpha P'$, on the other hand, is an exclusive feature of active polar systems and is ultimately responsible for the existence of traveling waves. 

For small values of $\alpha$ the solution of Eqs. \eqref{eq:simple} is the homogeneous state $(P,\omega)=(\pm \sqrt{a},0)$. A small perturbation of the homogeneous state of the form $\delta P(t) = P_{k}(t) e^{ikx}$ and $\delta \omega (t) = \omega_{k}(t) e^{ikx}$ evolves according to the equations:
\[
\frac{d}{dt}
\left(
\begin{array}{c}
\delta P_{k} \\
\delta \omega_{k}
\end{array}
\right)
= 
\left(
\begin{array}{cc}
-2a-k^{2}-ik\alpha & b \\
-\alpha k^{2} & -\eta k^{2}	
\end{array}
\right)
\left(
\begin{array}{c}
\delta P_{k} \\
\delta \omega_{k}
\end{array}
\right)\;.
\]
Thus, upon lowering $\alpha$ below the critical value $\alpha_{c}\sim-2a\eta/b$ (for small $k$) the homogeneous state becomes unstable and the solution consists of a wave train traveling along the negative $x$ direction at constant velocity. In order to describe the behavior of the traveling waves at the onset of the transition, we can construct an approximate solution of Eqs. \eqref{eq:simple} using the method of harmonic balance. We then look for a solution of \eqref{eq:simple} of the following form: 
\begin{gather*}
P(x,t) = P_{k}(t)e^{ikx}+\bar{P}_{k}(t)e^{-ikx}\;,\\[7pt] 
\omega(x,t) = \omega_{k}(t)^{ikx}+\bar{\omega}_{k}(t)e^{-ikx}\;,
\end{gather*}
with the bar indicating the complex conjugation. Substituting this in \eqref{eq:simple} and taking only the lowest wave numbers yields the following system of ordinary differential equations for the amplitudes:
\begin{subequations}\label{eq:amplitudes}
\begin{align}
\dot{P}_{k} &= - (k^{2}-a+ik\alpha)P_{k}-3\bar{P}_{k}P_{k}^{2}+b\omega_{k}\;, \\[5pt]
\dot{\bar{P}}_{k} &= - (k^{2}-a-ik\alpha)P_{k}-3P_{k}\bar{P}_{k}^{2}+b\bar{\omega}_{k}\;, \\[5pt]
\dot{\omega}_{k} &= -k^{2}(\eta\,\omega_{k}+\alpha P_{k})\;,\\[5pt]
\dot{\bar{\omega}}_{k} &= -k^{2}(\eta\,\bar{\omega}_{k}+\alpha \bar{P}_{k})\;.
\end{align}
\end{subequations}
These equations admits solutions of the form:
\[
P_{k}(t) = |P_{k}|\,e^{i(\nu t+\phi_{1})}\,,\qquad
\omega_{k}(t) = |\omega_{k}|\,e^{i(\nu t+\phi_{2})}\,.
\]
\begin{figure}[t]
\centering
\includegraphics[width=0.8\columnwidth]{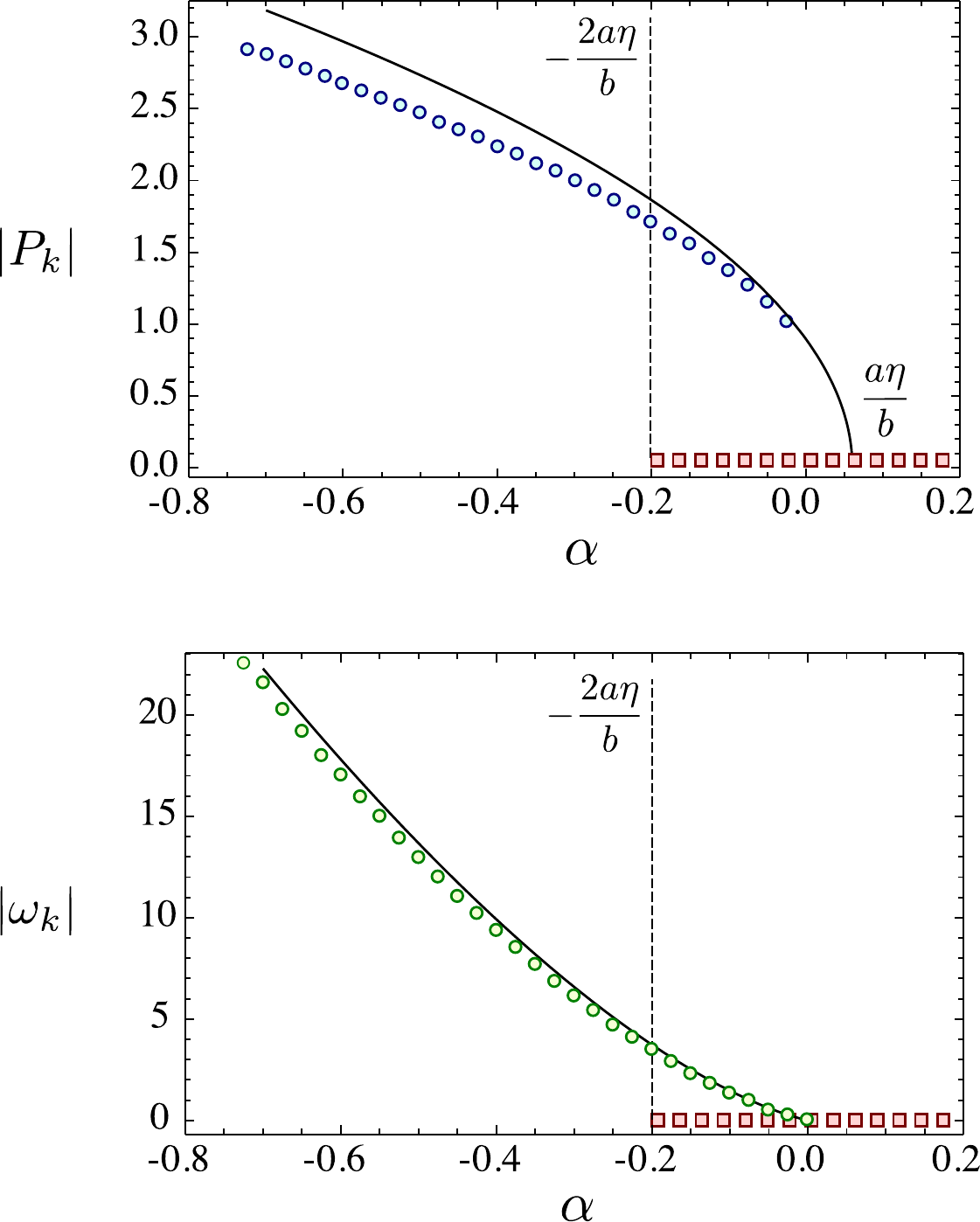}
\caption{\label{fig:hopf}(color online) The amplitudes of the traveling waves resulting from the numerical solution of the simplified model \eqref{eq:simple} together with the predictions of the harmonic balance analysis \eqref{eq:traveling-waves} for the largest spatial mode $k=2\pi/L$ (solid lines). At $\alpha\approx-2a\eta/b$ the system undergoes a subcritical Hopf bifurcation with a bistability region in the range $-2a\eta/b<\alpha<0$. The  red squares and blue/green circles correspond respectively to the stable homogeneous steady state and the stable large-amplitude limit cycle.}
\end{figure}
Without loss of generality we can take $\phi_{1}=0$ and $\phi_{2}=\phi$. Then, substituting this form in \eqref{eq:amplitudes} we find, after some manipulations, the following set of algebraic equations for the quantities $\nu$, $\phi$, $|P_{k}|$ and $|\omega_{k}|$:
\begin{subequations}
\begin{align}
&3|P_{k}|^{3}+(k^{2}-a)|P_{k}|-b|\omega_{k}|\cos\phi = 0\;, \\[7pt]
&\alpha k^{2} |P_{k}|+(\eta k^{2}\cos\phi-\nu\sin\phi)|\omega_{k}| = 0\;, \\[7pt]
&(k\alpha+\nu)|P_{k}|-b|\omega_{k}|\sin\phi = 0\;, \\[7pt]
&\nu \cos\phi+\eta k^{2}\sin\phi = 0\;.
\end{align}
\end{subequations}
Solving this equations to leading order in $\alpha$ gives:
\begin{gather}\label{eq:traveling-waves}
\nu \sim -k\alpha\,,\qquad
\phi \sim \frac{\alpha}{k\eta}\,,\\[5pt]
|P_{k}| \sim \sqrt{\frac{\eta (a-k^{2})-b \alpha}{3\eta}}\,,\quad
|\omega_{k}| \sim - \alpha\sqrt{\frac{\eta (a-k^{2})-b \alpha}{3\eta^{3}}}\,.\notag
\end{gather}
The larger spatial mode $k=2\pi/L$ thus evolves in time through traveling waves of the form:
\begin{gather*}
P(x,t) \approx 2|P_{k}|\cos k(x-\alpha t) \\[5pt]
\omega(x,t) \approx 2|\omega_{k}| \cos k(x-\alpha t -\alpha/\eta k^{2})
\end{gather*}
Fig. \ref{fig:hopf} shows a plot of the amplitudes $|P_{k}|$ and $|\omega_{k}|$ for the largest wavelength corresponding to $k=2\pi/L$ in Eqs. \eqref{eq:traveling-waves} together with those obtained from a numerical solution of the simplified equations \eqref{eq:simple}. 

The combined linear stability and harmonic balance analysis show that the Hopf bifurcation that leads to the formation of traveling waves is subcritical with bistable region in the range $-2a\eta/b<\alpha<0$. As the activity parameter $\alpha$ is lowered below zero, the system goes through a subcritical bifurcation where there is a stable stationary state and a stable large-amplitude limit cycle with an unstable periodic orbit acting as the separatrix between the two states. The existence of bistability at small negative values of $\alpha$ could have significant effects in the collective motion of active polar suspensions such as the presence of hysteresis when the amount of activity is cycled across the bifurcation point.

\subsection{Traveling Vortices and Chaos}

\begin{figure}
\centering
\includegraphics[width=1\columnwidth]{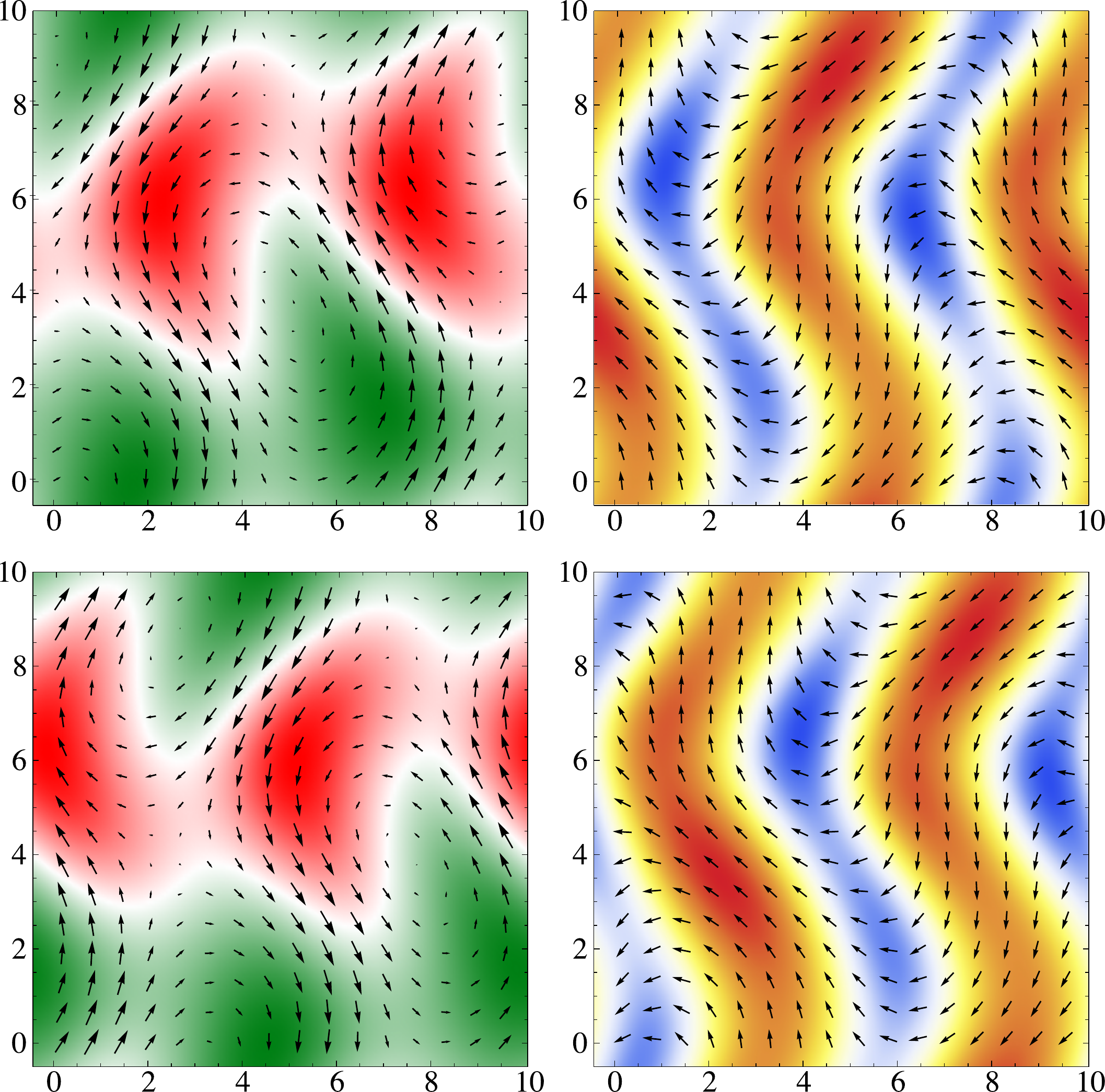}
\caption{\label{fig:traveling-vortices}(color online) The velocity field (left) and the polarization direction (right) superimposed to a density plot of the concentration and the polarization magnitude at two different times for $\alpha=-5$. The flow is characterized by two vortices and two stagnation saddle points traveling across the system from left to right.}	
\end{figure}

\begin{figure}
\centering
\includegraphics[width=1\columnwidth]{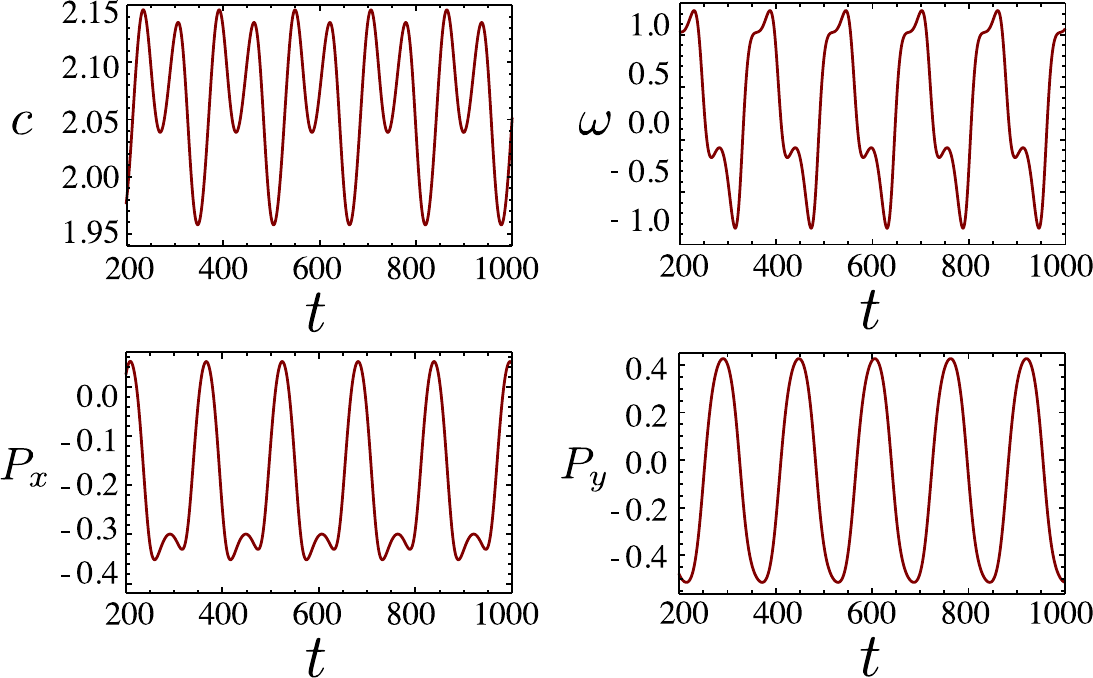}	
\caption{\label{fig:trajectory2}(color online) The quantities $c$, $\omega$, $P_{x}$ and $P_{y}$ at the point $x=y=L/2$ as a function of time in the traveling vortices for $\alpha=-5$.}
\end{figure}

If we further decrease $\alpha$, the traveling waves described in the previous section become unstable. In the new regime, the concentration is distributed in the form of kidney-shaped clusters that travels along the positive $x$ direction at constant speed (Fig. \ref{fig:traveling-vortices} right). The flow field, in turn, consists of a pair of oppositely spinning vortices whose centers also move at constant speed following the concentration clusters as well as two saddle stagnation points due to the toroidal topology of the domain  (Fig. \ref{fig:traveling-vortices} left). The order parameter appears concentrated along ``valleys'' whose curved conformation partially resembles the banded structure of the previous solution. Once again the relative variation in polarization is larger than that in concentration so that the value of the order parameter at the bottom of the valleys is almost 50\% lower than along the ridges. 

The formation of the vortices is associated with a break-down of the translational invariance along the longitudinal direction of the bands. At the onset of vortex formation the vorticity function for the spatial mode of largest wave-length is given heuristically by: 
\begin{equation}
\omega_{k}({\bf x},t) \approx \omega_{\bot}\cos k(x-\alpha P_{x}t)-\omega_{\|}\cos ky
\end{equation}
whose corresponding velocity field consists of two vortices and two saddles traveling along the $x$ direction with velocity $\alpha P_{x}$. From the numerical data shown in Fig. \ref{fig:traveling-vortices} we see that $P_{x}$ is everywhere negative, hence the vortices move toward the positive $x$ direction. In contrast with the traveling waves, where the polarization vector oscillate around its initial orientation (${\bf\hat{x}}$ in this case), the formation of traveling vortices is anticipated by an inversion of the polarization vector (i.e. ${\bf p}\sim -{\bf\hat{x}}$). The origin of this inversion is likely due to the short time scale at which the active stress is initially injected in the system, which makes the polarization field undergo dramatic distortions before it is able to settle on a limit cycle. 

\begin{figure}
\centering
\includegraphics[width=0.8\columnwidth]{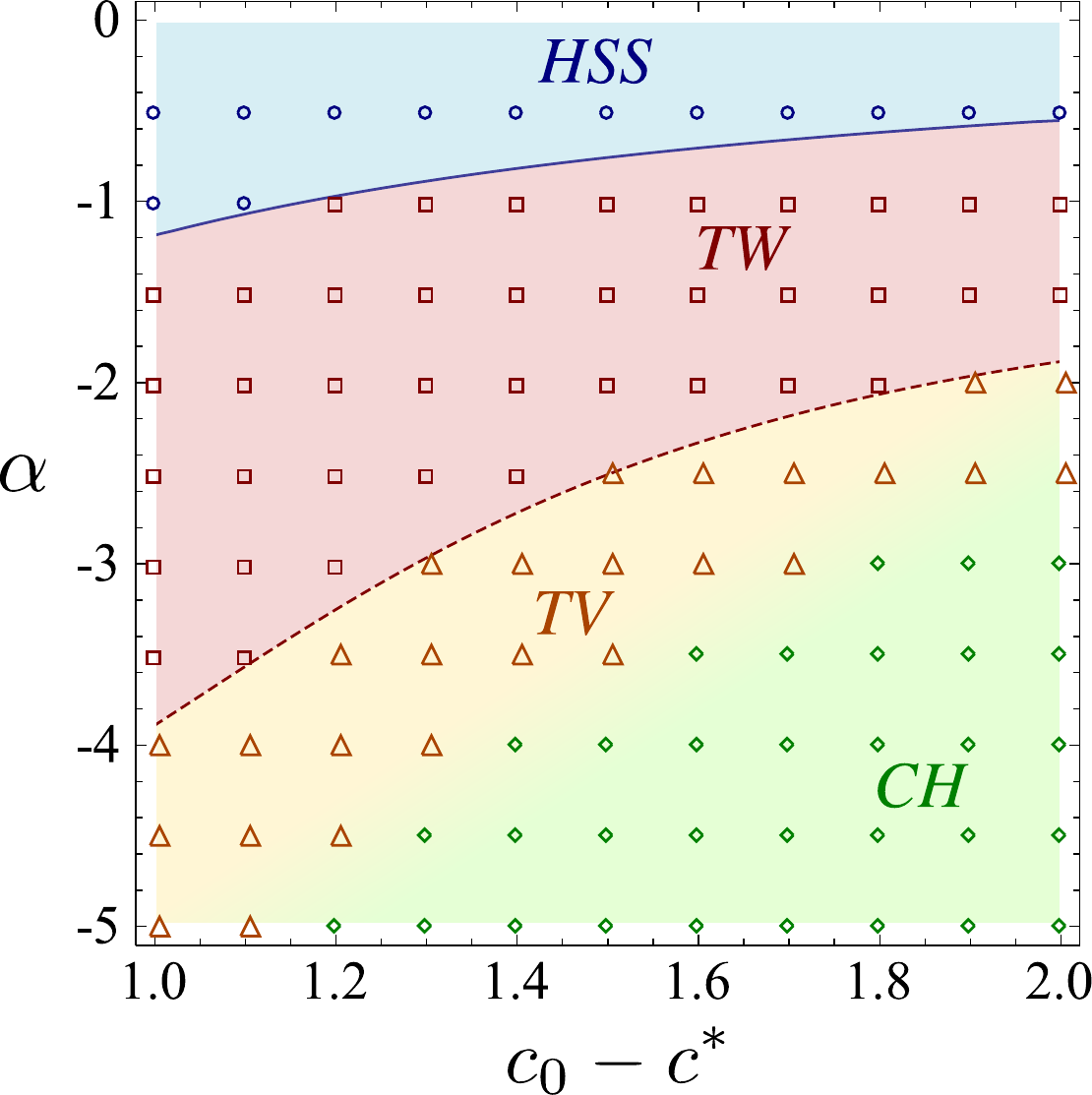}
\caption{\label{fig:nonlinear-pd}(color online) Phase diagram of the nonlinear spatiotemporal patterns: homogeneous steady state ({\em HSS}), traveling waves ({\em TW}), traveling vortices ({\em TV}) and chaos ({\em CH}). The dots have been obtained from the numerical integration of Eqs. \eqref{c-eq}, \eqref{P-eq} and \eqref{NS} for $\lambda=0.1$, $\eta=D_{0}=D_{1}=1$, $L=10$ and $w=0.1\alpha$. The solid blue line separating the homogeneous steady state from the traveling waves regime is given by Eq. \eqref{eq:alpha10}. The dashed line separating the traveling waves and the traveling vortices regime has been inferred from the data. The color gradient between the traveling vortices and the chaotic regime indicates a fuzzy boundary line.}	
\end{figure}

Upon further decreasing $\alpha$, the traveling vortices become unstable and the system undergoes a transition to a chaotic regime. The polarization valleys now continuously form and disappear in unpredictable fashion. The flow still exhibits larger vortices, which however move chaotically across the sample with variable direction and magnitude. Interestingly, the polarization direction is continuously distorted and does not posses regions of partial alignment. This differs from the spatio-temporal chaos found active nematics \cite{Giomi2011} where the chaotic flow is organized in grains of uniform orientation separated by chaotically moving grain boundaries. \revision{The nonlinear spatiotemporal patterns described in this section are summarized in the phase-diagram of Fig. \ref{fig:nonlinear-pd}. The Vicsek-type traveling waves shown in the linear phase diagram of Fig.~\ref{fig:phase-diagram} are also obtained from the numerical solution of the nonlinear equations, but at a much lower density that shown in Fig.~\ref{fig:nonlinear-pd}.}

\begin{figure}
\centering
\includegraphics[width=1\columnwidth]{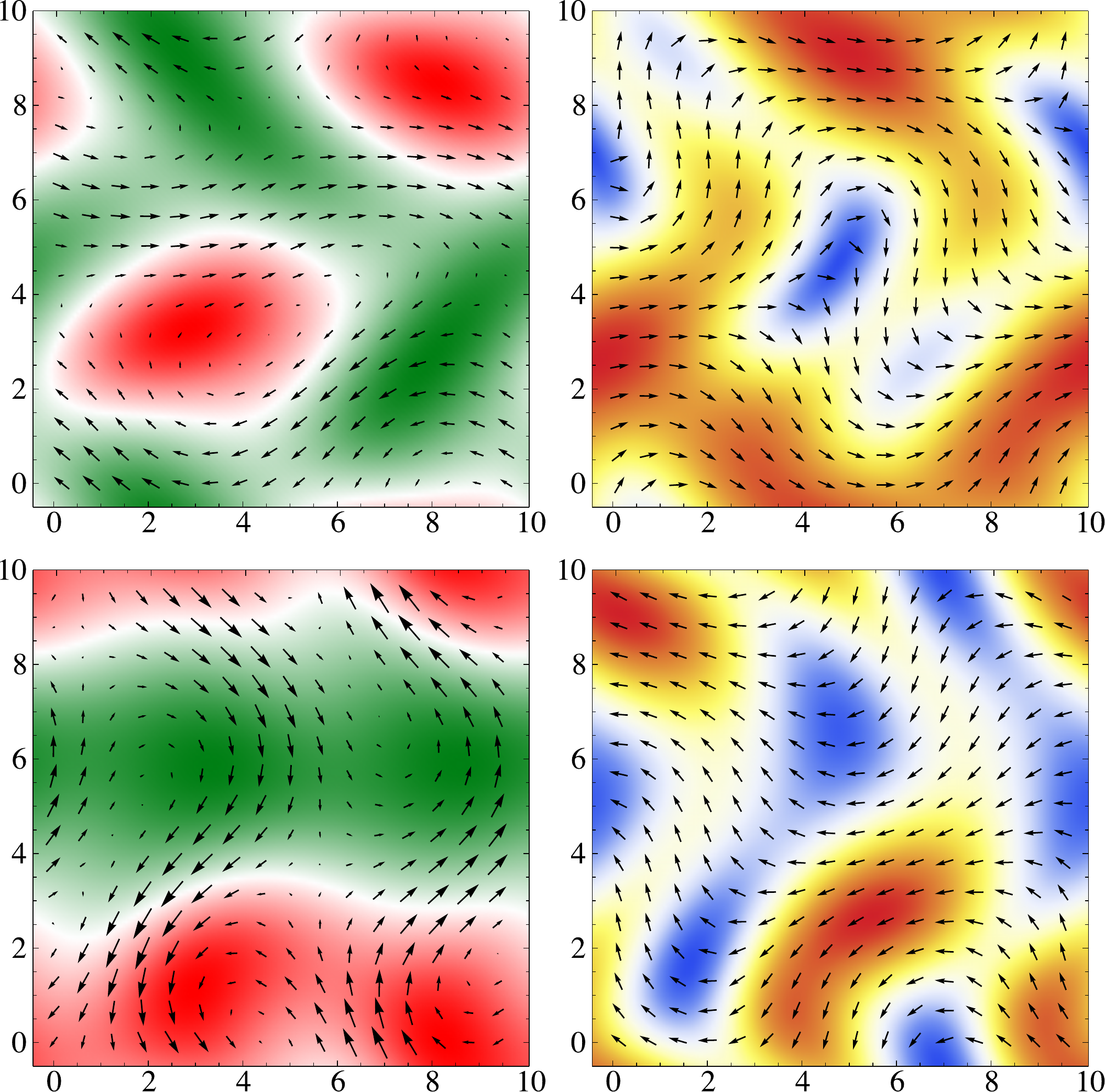}
\caption{\label{fig:chaos}(color online) The velocity field (left) and the polarization direction (right) superimposed to a density plot of the concentration and the polarization magnitude for $\alpha=-6$. The flow consists of large vortices that move chaotically across the system with variable direction and magnitude.}	
\end{figure}

\section{\label{sec:Discussion}Discussion}

We have analyzed the complex spatio-temporal patterns of an active polar fluid in two dimensions. The continuum model considered is appropriate to describe systems such as bacterial suspensions, consisting of self-propelled particles in a solvent. The self-propulsion of the active units yields a number of advective terms in the hydrodynamic equations that are unique to polar fluids. These advective terms are responsible for the traveling nature of the large scale structures, which include traveling stripes oriented normal to the direction of propagation and traveling vortices.  These structures are remarkably stable in a wide region of parameters and resemble the patterns seen recently in actin motility assays at high actin density, suggesting that systems may indeed have an underlying polar symmetry. We stress that the transitions between regimes characterized by different stable large-scale structures can only be obtained by allowing for spatial variations of the order parameter.  A novel feature of the present work is that the transitions between these various regimes appear to be discontinuous, allowing for the possibility of coexistence and hysteresis. Discontinuous transitions between disordered and and ordered states have also been seen in dry active systems via numerical studies of the Vicsek model~\cite{Chate2008}. Understanding the nature and order of the transition in both suspensions and active particles on substrates will, however, require further investigation.

\revision{The present analysis is limited to two dimensional systems and may be applicable to situations like motility assays where the system is quasi-2D. One should, however, be cautious about generalizing the results found in 2D to 3D systems, as new patterns may arise one the polarization field is allowed to rotate out of the plane. Very recent work by Fielding et al.~\cite{Fielding2011} has indeed shown that the behavior of a sheared active nematic in two dimensions can be quite different from that in one dimension.}

Our work, together with recent work on pattern formation inactive nematic, suggests a strong analogy between the spatio-temporal dynamics of active fluids and the behavior of excitable systems. In both cases the complex behavior obtained from the nonlinear hydrodynamic equations can be reproduced qualitatively by simplified lower dimensional models that resemble reaction-diffusion or reaction-diffusion-advection equations familiar from the study of excitable systems. This mapping may provide a useful framework for understanding and classifying active systems. 

\acknowledgments

MCM was supported by the NSF on grants DMR-0806511 and DMR-1004789. LG was supported by the Harvard-NSF MRSEC, the Harvard-Kavli Nano-Bio Science and Technology Center and the Wyss Institute. We thank Aparna Baskaran and Paolo Paoletti for illuminating discussions.

\appendix

\section{\label{sec:AppendixA}Appendix A}
In this appendix we give the explicit expressions for the elements of the matrix ${\bf A}_{nm}$ given in Eq.~\eqref{Anm}. The various block elements are given by
\begin{widetext}
\begin{equation}
\label{a-nm}
\bm{a}_{nm} =
\left(
\begin{array}{cc}
- D\left(q_m^2+q_n^2\right)  -i w q_n P_0 &
-i w q_n c_0 \\
-(a'_2+a'_4P_0^2)P_0-\frac{i wq_n}{2c_0} & 
- K\left(q_m^2+q_n^2\right)-i w q_n P_0 -2 |a_{20}|
\end{array}
\right)
\end{equation}
\[
\bm{b}_{nm}
\left(
\begin{array}{cc}
-i w q_m c_0  & 0 \\
 0 & -\frac{q_m q_n \lambda  P_0}{q_m^2+q_n^2}
\end{array}
\right)
\]
\[
\bm{c}_{nm} =
\left(
\begin{array}{cc}
-\frac{i wq_m }{2 \mathit{c}_0} 
& 0 \\[10pt]
C_{nm}^{21} &  
C_{nm}^{22}
\end{array}
\right)
\]
with:
\[
C_{nm}^{21} = \alpha  P_0^2\left(q_m^2+q_n q_m-q_n^2\right)-\frac{2q_ m q_n \lambda  c^{*} P_0^2}{c_0+c^{*}}-\frac{iq_m w P_0  (1-\lambda ) \left(q_m^2+q_n^2\right)}{4 c_0}
\]
\[
C_{nm}^{22} = 
2 c_0 P_0 \left[q_m q_n \lambda +\alpha \left(q_m^2+q_n q_m-q_n^2\right) \right]
+q_m q_n  \lambda  P_0 \left[\left(q_m^2+q_n^2\right) -2 c^{*}\right]
\]
and:
\[
\bm{d}_{nm}
= \left(
\begin{array}{cc}
 - \left[  \left(q_m^2+q_n^2\right)+i q_n  w  P_0\right]& \frac{P_0\left[(1-\lambda ) q_m^2+q_n^2 (\lambda +1)\right]}{2 \left(q_m^2+q_n^2\right)} \\[10pt]
 c_0 \alpha  P_0 \left(q_m^2-q_n^2\right)-\frac{  P_0}{2} \left(q_m^2+q_n^2\right) \left[(1-\lambda ) q_m^2+q_n^2 (\lambda
   +1)\right] & - \eta \left(q_m^2+q_n^2\right)
\end{array}
\right)
\]
where $a_{20}=a_2(c_0)=c^*-c_0<0$, $D_{0}=D_{1}=D$ and the prime denotes a derivative with respect to concentration, i.e., $a'_{2,4}=\left(\partial a_{2.4}/\partial c\right)_{c=c_0}$, which gives $-(a'_2+a'_4P_0^2)=2c^*/(c_0+c^*)$.
\end{widetext}

\section{\label{sec:AppendixB}Appendix B}

Analytical expression for the values of $\alpha$ associated with these instabilities can be found by diagonalizing the matrix $\bm{d}_{nm}$. For both the longitudinal (10) and the transverse (01) directions, it is the coupled modes of $P_{y}$ and $\omega$ that go unstable. For $\alpha>0$ the modes go unstable along the 01 direction in this case $P_{y}$ corresponds to splay deformations and the vorticity describes fluctuations in the $v_{x}$ velocity. For $\alpha<0$ the modes go unstable along the 10 direction in this case $P_{y}$ corresponds to bend deformations and the vorticity describes fluctuations in the $v_{y}$ velocity. Both these instabilities exist also when $w=0$. The matrices $\bm{d}_{10}$ and $\bm{d}_{01}$ are given by:
\[
\bm{d}_{10} = 
\left(
\begin{array}{cc}
	
	-q^{2}-iqwP_{0}		&	\frac{1}{2}(1+\lambda)P_{0} \\[5pt]
	-q^{2}P_{0}[\frac{1}{2}q^{2}(1+\lambda)+\alpha c_{0}] & -\eta q^{2}
	
\end{array}
\right)
\]
and:
\[
\bm{d}_{01} = 
\left(
\begin{array}{cc}
	
	-q^{2}	&	\frac{1}{2}(1-\lambda)P_{0} \\[5pt]
	-q^{2}P_{0}[\frac{1}{2}q^{2}(1-\lambda)-\alpha c_{0}] & -\eta q^{2}
	
\end{array}
\right)
\]
with $q=2\pi/L$. The corresponding eigenvalues are:
\begin{widetext}
\begin{gather*}
\Lambda_{10}
= -\frac{1}{2}q\left\{q(\eta+1)+iwP_{0}\pm \sqrt{q^{2}(\eta-1)^{2}-P_{0}^{2}[w^{2}+q^{2}(1+\lambda)^{2}+\alpha c_{0}(1+\lambda)] -2 i q w P_{0}(\eta-1)} \right\}\\[5pt]
\Lambda_{01} = -\frac{1}{2}q\left\{q(\eta+1)\pm\sqrt{q^{2}(\eta-1)^{2}+P_{0}^{2}(1-\lambda)[2\alpha c_{0}-q^{2}(1-\lambda)]}\right\}
\end{gather*}
\end{widetext}


\begin{thebibliography}{99}

\bibitem{Gruler1999}
H. Gruler, U. Dewald, and M. Eberhardt,
Eur. Phys. J. B {\bf 11}, 187 (1999). 

\bibitem{TonerTu1995}
J. Toner and Y. Tu, 
Phys. Rev. Lett. {\bf 75}, 4326  (1995).

\bibitem{TonerRev} 
J. Toner, Y. Tu and S. Ramaswamy, 
Ann. Phys. {\bf 318}, 170 (2005).

\bibitem{SimhaRamaswamySP02}
R. A. Simha and S. Ramaswamy, 
Phys. Rev. Lett. \textbf{89}, 058101 (2002).

\bibitem{EdwardsYeomans2009}
S. A. Edwards and J. M. Yeomans, Europhys. Lett.  {\bf 85}, 18008 (2009).

\bibitem{Kruse2004}  
K. Kruse, J. F. Joanny, F. J\"{u}licher, J. Prost, and K. Sekimoto,
Phys. Rev. Lett. {\bf 92}, 078101  (2004).

\bibitem{Kruse2005}  
K. Kruse, J. F. Joanny, F. J\"{u}licher, J. Prost, and K. Sekimoto,
Eur. Phys. J. E {\bf 16}, 5Ð16 (2005).

\bibitem{SaintillanShelley:2007}
D. Saintillan and M. J. Shelley,
Phys. Rev. Lett., {\bf 99} 058102 (2007).

\bibitem{SaintillanShelley:2008}
D. Saintillan and M. J. Shelley,
Phys. Rev. Lett., {\bf 100} 178103 (2008).

\bibitem{TBLMCM2003}  
T. B. Liverpool and M. C. Marchetti,
Phys. Rev. Lett. {\bf 90}, 138102 (2003).

\bibitem{TBLMCMbook}
T. B. Liverpool and M. C. Marchetti, in \emph{Cell Motility}, 
P. Lenz, ed. (Springer, New York, 2007).

\bibitem{Voituriez06}
R. Voituriez, J. F. Joanny and J. Prost, 
Europhys. Lett. \textbf{70}, 118102 (2005).

\bibitem{Marenduzzo2007}
D. Marenduzzo, E. Orlandini, M. E. Cates and J. M. Yeomans, Phys. 
Rev. E {\bf 76}, 031921 (2007).

\bibitem{GiomiMarchettiLiverpool:2008}
L. Giomi, M. C. Marchetti and T. B. Liverpool, Phys. Rev. Lett. {\bf 101}, 198101 (2008).

\bibitem{Hatwalne04}
Y. Hatwalne, S. Ramaswamy, M. Rao and R. A. Simha,
Phys. Rev. Lett. \textbf{92}, 118101 (2004)

\bibitem{TBLMCM06}
T. B. Liverpool,M. C. Marchetti, 
Phys. Rev. Lett. \textbf{97}, 268101 (2006)

\bibitem{CatesEtAl:2008}
M. E. Cates, S. M. Fielding, D. Marenduzzo, E. Orlandini and J. M. Yeomans, 
Phys. Rev. Lett. {\bf 101}, 068102 (2008).

\bibitem{GiomiTBLMCM-shear2010}
L. Giomi, T. B. Liverpool and M. C.
Marchetti, Phys. Rev. E {\bf 81}, 051908 (2010). 

\bibitem{Saintillan2010}
D. Saintillan,
Phys. Rev. E {\bf 81}, 056307 (2010).

\bibitem{Rafai2010}
S. Rafa\"i, L. Jibitu and P. Peyla,
Phys. Rev. Lett. {\bf 104}, 098102 (2010).

\bibitem{SokolovAranson2009}
A. Sokolov and I. S. Aranson,
Phys. Rev. Lett. {\bf 103}, 148101 (2009).

\bibitem{Giomi2011}
L. Giomi, L. Mahadevan, B. Chakraborty and M. F. Hagan,
Phys. Rev. Lett. {\bf 106}, 218101 (2011). 

\bibitem{Fielding2011}
S. M. Fielding, D. Marenduzzo and M. E. Cates,
Phys. Rev. E {\bf 83}, 041910 (2011).

\bibitem{Schaller2010}
V. Schaller, C. Weber, C. Semmrich, E. Frey and A. R. Bausch, 	
Nature {\bf 467}, 73 (2010).

\bibitem{Butt2010}
T. Butt, T. Mufti, A. Humayun, P. B. Rosenthal, S. Khan, S. Khan and J. E. Molloy,
J. Biol. Chem. {\bf 285}, 4964 (2010).

\bibitem{Kaiser2003}
D. Kaiser, 
Nat. Rev. Microbiol. {\bf 1}, 45 (2003).

\bibitem{Zumsan2007}
D. R. Zusman, A. E. Scott , Z. Yang and  J. R. Kirby,
Nat. Rev. Microbiol. {\bf 5}, 862 (2007).

\bibitem{Peruani2006}
F. Peruani, A. Deutsch, and M. B\"ar, 
Phys. Rev. E {\bf 74}, 030904(R) (2006).

\bibitem{Chate2008}
H. Chat\'e, F. Ginelli, G. Gr\'egoire and F. Raynaud,
Phys. Rev. E {\bf 77}, 046113 (2008).

\bibitem{GinelliPeruani2010}
F. Ginelli, F. Peruani, M. B\"ar and H. Chat\'e,
Phys. Rev. Lett.  {\bf 104}, 184502 (2010).

\bibitem{Yang2010}
Y. Yang, V. Marceau and G. Gompper,	
Phys. Rev. E {\bf 82}, 031904 (2010).

\bibitem{BDG2009}
E. Bertin, M. Droz and G. Gr\'egoire, 
J. Phys. A: Math. Theor. {\bf 42 } 445001 (2009).

\bibitem{Ihle2011}
T. Ihle,
Phys. Rev. E {\bf 83}, 030901(R) (2011).

\bibitem{Mishra2010}
S. Mishra, A. Baskaran, and M. C. Marchetti,
Phys. Rev. E {\bf 81}, 061816 (2010).

\bibitem{Voituriez2006}
R. Voituriez, J. F. Joanny, and J. Prost, Phys. Rev. Lett. {\bf 96}, 028102 (2006).

\bibitem{Blankschtein1985}
D. Blankschtein and R. M. Hornreich, Phys. Rev. B {\bf 32}, 
3214 (1985).

\bibitem{defnote}
The  parameters $\beta$ and $\alpha$ have different dimensions  from those used in~\cite{TBLMCMbook} as here we have incorporated an additional factor of $\ell^2$ in their definition.

\bibitem{alpha-sign}
We use the sign convention introduced in Ref.~ \cite{TBLMCM2003}, which is opposite to that used on much later work in acto-myosin systems (e.g., Ref.~\cite{Kruse2005}).

\bibitem{BaskaranMarchetti:2009}
A. Baskaran and M. C. Marchetti,
Proc. Natl. Acad. Sci. USA {\bf 106}, 15567 (2009).

\bibitem{TonerTu-note}
The hydrodynamics of a dry polar active systems was first considered by Toner and Tu in Ref.~\cite{TonerTu1995}. The velocity used by these authors as the order parameter is simply proportional to our polarization ${\bf P}$ and should not be confused with the flow velocity introduced in the present work.

\bibitem{Mishra-splay-note}
A splay instability is obtained in Ref.~\cite{Mishra2010} by a different mechanism.

\bibitem{Turing:1952} 
A. M. Turing,
Phil. Trans. R. Soc. B {\bf 237}, 37 (1952).
	
\end{thebibliography}
\end{document}